# Long-Term Stabilization of Two-Dimensional Perovskites by Encapsulation with Hexagonal Boron Nitride

**Michael Seitz[1], Patricia Gant[2], Andres Castellanos-Gomez[2]\*, Ferry Prins[1]\***

1. Condensed Matter Physics Center (IFIMAC), Universidad Autónoma de Madrid, 28049 Madrid, Spain; ferry.prins@uam.es
2. Materials Science Factory, Instituto de Ciencia de Materiales de Madrid. Consejo Superior de Investigaciones Científicas, 28049 Madrid, Spain; andres.castellanos@csic.es
* Correspondence: ferry.prins@uam.es; andres.castellanos@csic.es

**Abstract:** Metal halide perovskites are known to suffer from rapid degradation, limiting their direct applicability. Here, the degradation of phenethylammonium lead iodide ($PEA_2PbI_4$) two-dimensional perovskites under ambient conditions is studied using fluorescence, absorbance and fluorescence lifetime measurements. It is demonstrated that a long-term stability of two-dimensional perovskites can be achieved through the encapsulation with hexagonal boron nitride. While un-encapsulated perovskite flakes degrade within hours, the encapsulated perovskites are stable for at least three months. In addition, encapsulation considerably improves the stability under laser irradiation. The environmental stability, combined with the improved durability under illumination, is a critical ingredient for thorough spectroscopic studies of the intrinsic opto-electronic properties of this material platform.

## 1. Introduction

Metal halide perovskites ($APbX_3$, A = methylammonium, formamidinium, Cs; X = I, Br, Cl), have recently emerged as a promising material class for light harvesting [1–4] and light emitting [5–11] applications. Most prominently, perovskite solar cells have reached light collection efficiencies of well over 20% just a few years after their first reports [4]. Rapid advances are made possible through an excellent combination of favorable properties such as solution processability at low temperatures [2,12], remarkable tolerance to defects [13–17], high absorption coefficients [2,12], long-range ambipolar charge transport characteristics [18–20], and the broad tunability of their bandgap, freely adjustable from the ultra-violet into the infra-red [21–23].

Despite these advantages, the study and applicability of metal halide perovskites remain challenging due to their poor environmental stability [3,24–26]. Perovskites normally degrade within several hours or days under ambient condition with oxygen, moisture, light irradiation, and heat playing an important role in the degradation process [3,24–26]. Findings of several studies on lead iodide perovskites suggest that during degradation the bulk perovskite $CH_3NH_3PbI_3$ decomposes into the volatile species $I_2$ and $CH_3NH_3$ (methylammonium) leaving behind $PbI_2$ [27–30]. The cause of this degradation is ascribed to oxygen [29,31,32], moisture [27–29,33,34], photoactivation [27,32,35,36], and/or heat [29,35]. [3,24,25]

One possible route to improve the stability of perovskites is through tuning of their chemical composition, for example by exchanging the methylammonium cation with the less volatile formamidinium [37], cesium [38], rubidium [39], or a mixture thereof [40–43]. Also, a partial or complete substitution of iodine with bromine [44,45], chlorine [46], or a mixture of the two [47] has been shown to yield more stable perovskites [26]. However, despite these efforts, the stability of the perovskite crystals remains limited. Another promising strategy to significantly improve the stability of perovskite devices is to reduce the dimensionality of the perovskite crystals [26]. A prominent example are layered two-dimensional (2D) perovskites, which have a higher formation energy than their bulk counterparts due to closely bound ligands which separate the inorganic layers. Noticeably, Grancini et al. [48] have shown that 2D/3D perovskite solar modules show a prolonged stability.

An alternative strategy to improve the stability of perovskites is through encapsulation, which prevents water and air molecules from reaching the perovskite crystal in the first place [24,25,49–53]. For example, by encapsulation of a complete photovoltaic device, Gevorgyan et al. [54] produced a perovskite solar cell which was stable for over a year. Further, it has been shown that 2D materials suffering from air-induced degradation, such as black phosphorus, can be protected by encapsulation with hexagonal boron nitride (hBN) [55,56]. Indeed, recent studies have shown this hBN encapsulation on such a smaller level is also possible for perovskites, providing improved resistance to heat, moisture, and light irradiation [30,57–59].

Here, we study the feasibility of the hBN encapsulation strategy to provide long-term stability of phenethylammonium lead iodide ($PEA_2PbI_4$) 2D perovskites. While unprotected flakes degrade within hours, hBN encapsulated perovskite flakes display no observable degradation in their optical properties for at least three months. Further, we report a significantly improved resistance towards laser radiation. The environmental stability, combined with the improved durability under illumination, is a critical ingredient for thorough spectroscopic studies of the intrinsic opto-electronic properties of this material platform. This is of particular importance for thin flakes, where degradation effects are more pronounced.

## 2. Materials and Methods

*Chemicals:* Chemicals were purchased from commercial suppliers and used as received: lead(II) iodide $PbI_2$ (Sigma Aldrich 900168-5G), phenethylammonium iodide PEAI (Sigma Aldrich 805904-25G), γ-butyrolactone (Sigma Aldrich B103608-500G), hexagonal boron nitride hBN (HQ Graphene).

*2D Perovskite Synthesis:* Layered perovskites were synthesized under ambient laboratory conditions following the over-saturation techniques reported previously [60,61]. In a nutshell, the $PbI_2$ (200 mg) and PEAI (216 mg) were mixed in a 1:2 molar ratio and dissolved in γ-butyrolactone (300 μl). The solution was heated to 70 °C and more γ-butyrolactone (0-200 μl) was added until all the precursors were completely dissolved. After 2-3 days, millimeter sized crystals formed in the solution, which was subsequently cooled down to room temperature. For this study, we dropcast some of the remaining supersaturated solution on a glass slide (the millimeter sized crystals were used for another study), heated it up to 70 °C with a hotplate and after the solvent was evaporated, $PEA_2PbI_4$ crystals with crystal sizes of up to several hundred microns were formed. The saturated solution can be stored and re-used to produce freshly grown 2D perovskites within several minutes.

*Exfoliation:* The synthesized $PEA_2PbI_4$ perovskite crystals were mechanically exfoliated using the Scotch tape method (Nitto SPV 224). After an appropriate flake thickness was obtained, the flakes were exfoliated one last time with polydimethylsiloxane (PDMS, Gelfilm from Gelpak®), which is better suited for the precise placement of the flakes. Following this method, we were able to obtain thin single crystals with sizes up to hundreds of microns. After several exfoliations between two scotch tapes and exfoliation with the PDMS, we transferred the flakes on a glass substrate. We found that retracting the tape and PDMS stamp quickly transfers thick and large crystals, while retracting slowly yields thinner crystals.

*Characterization:* Fluorescence and absorption spectra were recorded using a spectrograph and an EMCCD camera from Princeton Instruments (SpectraPro HRS-300, ProEM HS 1024BX3) with a 300 g/mm grating with a blaze of 500 nm. The samples were excited by a LED (Thorlabs M385PLP1-C5, λ = 385 nm) for fluorescence measurements (0.55 mWcm$^{-2}$, 1 min exposure) and by a white light LED (Thorlabs MCWHLP1) for absorbance measurements (0.12 mWcm$^{-2}$, 1 min exposure). Fluorescence lifetime measurements were performed with a laser diode of λ = 405nm (PicoQuant LDH-D-C-405, PDL 800-D, Pico-Harp 300) and an avalanche photo diode (APD) from micro photon devices (Micro Photon Devices PDM). The repetition rate was 1 MHz and the peak fluence per pulse was 1 nJcm$^{-2}$. X-ray diffraction (XRD) was performed using a Bruker D8 Advance operating at 40 kV and 30 mA using a copper radiation source (1.54060 Å). XRD measurements were taken from the dropcast perovskite films.

## 3. Results and Discussion

2D perovskites PEA$_2$PbI$_4$ are synthesized by a one-step drop casting from a saturated precursor solution. The crystalline integrity of the perovskite film is confirmed by performing x-ray diffraction (XRD) analysis. As shown in **Figure S1**, the XRD measurement reveals evenly spaced peaks, which confirm the n = 1 perovskite structure with a spacing of 1.63 nm between the inorganic layers, consistent with previously reported values [62–64]. Through mechanical exfoliation of the crystalline perovskite film, we obtain 10 – 200 µm large single crystalline multilayer 2D flakes as shown in **Figure 1** [65]. Fluorescence and absorbance spectroscopy on the flakes reveal the characteristic green fluorescence at 527 nm known for PEA$_2$PbI$_4$ (**Figure S2**).

Only minutes after their exfoliation, the 2D perovskite flakes start to degrade into transparent and non-emissive PbI$_2$ [27–30]. The degradation starts at the surface and progresses into the flake as can be seen from the transmission and fluorescence images in Figure 1. The lateral degradation of the flake is faster. We assign this to the easier diffusion of water, oxygen and reaction products along the inorganic layers as compared to the layer-by-layer diffusion through the inorganic layers. However, the decrease of fluorescence and absorbance in the center of the flake, well before the lateral degradation reaches the center of the flake, suggests that layer-by-layer degradation is also present, although slower than the lateral degradation along the inorganic layers. It is important to notice that in thin films, the lateral degradation is hindered through neighboring crystallites, making it harder for air molecules to penetrate the perovskite film and thereby slowing down the overall degradation process. As a result, the degradation time of perovskites may vary for different studies, since the speed of the degradation is thickness and lateral-size dependent.

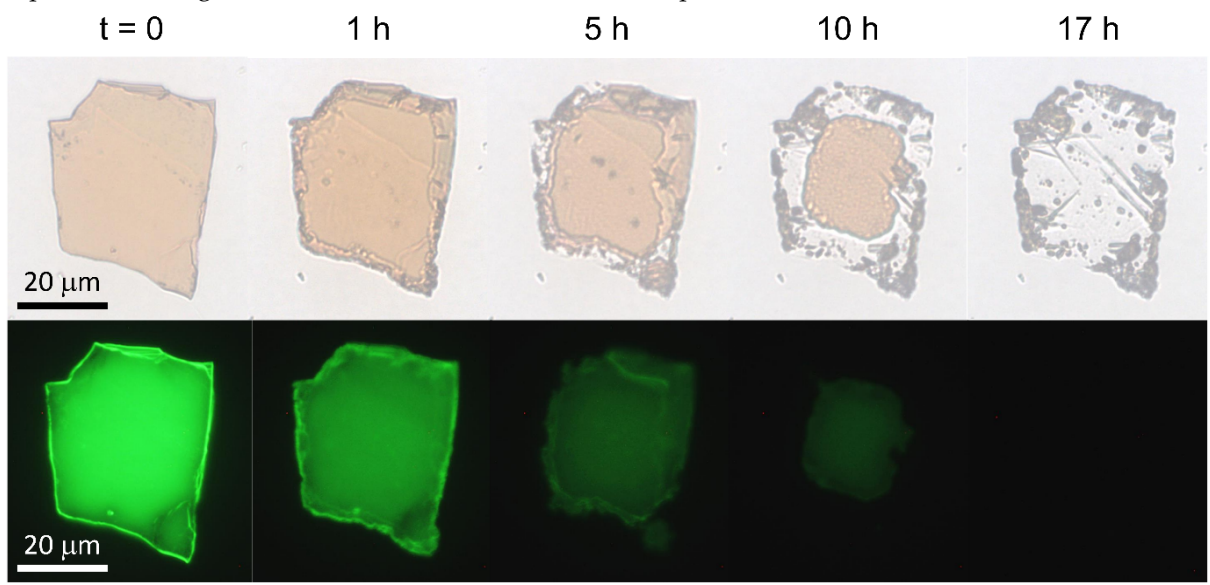

**Figure 1.** Complete degradation of an exfoliated PEA$_2$PbI$_4$ 2D perovskite flake within 17h of ambient conditions. First and second row show transmission and fluorescence micrographs of the flake, respectively.

To study the time-dependent degradation of 2D perovskite flakes, we monitor the fluorescence, absorbance and fluorescence lifetime of the flakes in more detail (**Figure 2**). During the measurements, exposure to light is minimized to avoid light-induced degradation as much as possible. As can be seen in Figure 2b, the degradation is most prominently visible in the fluorescence measurements. After a first initial drop, which we attribute to the formation of non-radiative trap states at the surface, a slower decay of fluorescence intensity (between 15 min and 5 h) is observed. This could be explained by a diffusion limited progression of degradation into the perovskite flake, during which the fluorescence drops exponentially (see also **Figure S3**). In addition to the decreasing fluorescence intensity, we observe a 4 nm blueshift of the emission peak. Likewise, we observe a decrease in the first absorption peak (Figure 2c,d) and the fluorescence lifetime (Figure 2e,f) although less rapidly than for the fluorescence. The faster degradation of the fluorescence as compared to the absorptive properties and the simultaneous decrease in fluorescence lifetime suggest the formation of additional non-radiative decay channels. We would like to note that the complete degradation depicted in Figure 2 is faster than the one in Figure 1, as the latter flake is thicker and has a larger lateral size, increasing the time needed for a complete degradation.

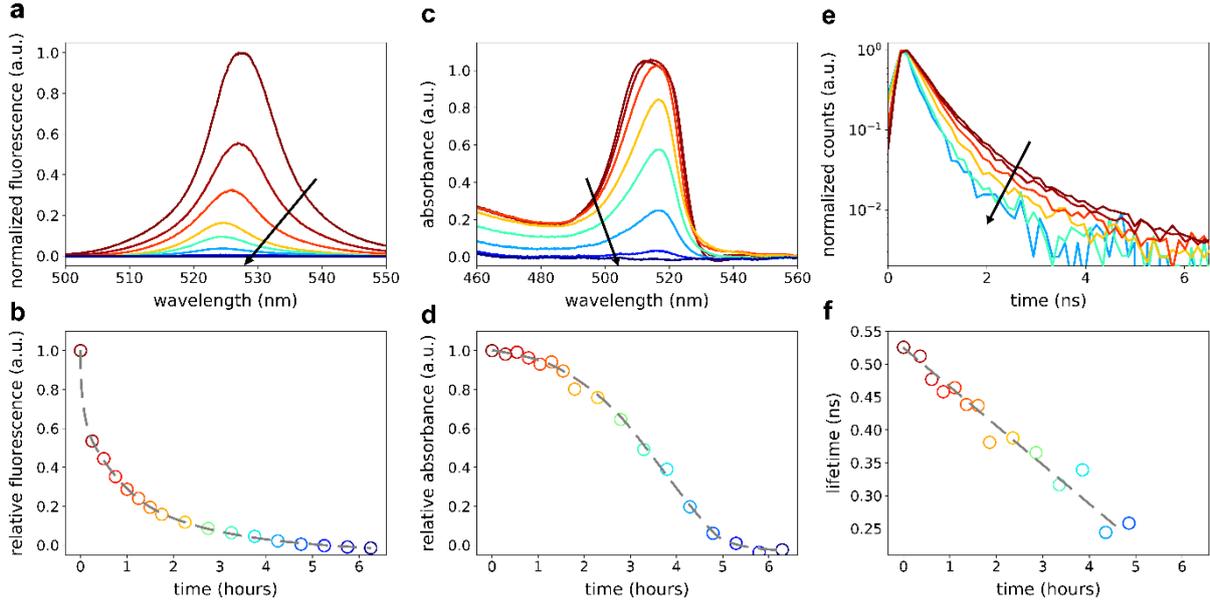

**Figure 2.** Spectral properties of a $PEA_2PbI_4$ 2D perovskite flake for different times under ambient exposure. a, c, e) Fluorescence, absorbance and fluorescence lifetime traces, respectively, for different times of ambient exposure. Traces shown correspond to times 0, 0.25, 1, 2.25, 3.25, 4.25, 5.25, and 6.25 h. Fluorescence lifetime traces are only shown for t < 5 h as the data is too noisy for later times. b) Total fluorescence intensity from (a) normalized to the measured intensity at t = 0. d) Integrated absorbance (455 nm – 594 nm) normalized to the measured absorbance at t = 0. f) Extracted 1/e lifetime from the fluorescence lifetime traces in (e). Gray lines in b), d), and f) are guides to the eye.

To prevent the rapid degradation, we fully encapsulate perovskite flakes between two hBN layers. Samples are prepared by mechanical exfoliation with a stamp of polydimethylsiloxane (Gelfilm from Gelpak®) from hBN crystals, which are transferred on a glass slide. In a next step, a perovskite flake, obtained by the same method, is placed onto the first hBN flake using an all dry deterministic placement method and finally the perovskite is encapsulated with a second mechanically exfoliated hBN flake (exposure of the perovskite flake to ambient conditions is < 10 min) using the same deterministic placement technique [66]. In **Figure 3a** we show a transmission microscopy image of the stack with dotted lines identifying the different layers.

Following the same procedures as for the un-encapsulated perovskite flake, we monitor the fluorescence, absorbance and fluorescence lifetime of the encapsulated flake. After three months, no significant degradation is observed on the encapsulated flake. This can be seen qualitatively in Figure 3a and Figure 3b-d, where we show that an encapsulated flake maintains its optical properties over the course of three months, while an un-encapsulated flake degrades within 5 hours. For the encapsulated flake the fluorescence and the fluorescence lifetime even increase slightly. We attribute this increase in fluorescence to photobrightening through a photo-induced halide redistribution that occurs during our spectral measurements, despite the low fluences during our measurements [67]. The fluorescence, absorbance and fluorescence lifetime traces of the encapsulated flake are shown in **Figure S4**. **Figure S5** shows a partially encapsulated flake with a small leak in the hBN seal. It is visible that the degradation starts at the leak and slowly progresses into the flake. However, the degradation is significantly slowed down because the diffusion of the air molecules is limited by the small hole in the hBN seal. Even after one month the center of the flake is not yet significantly affected by the degradation.

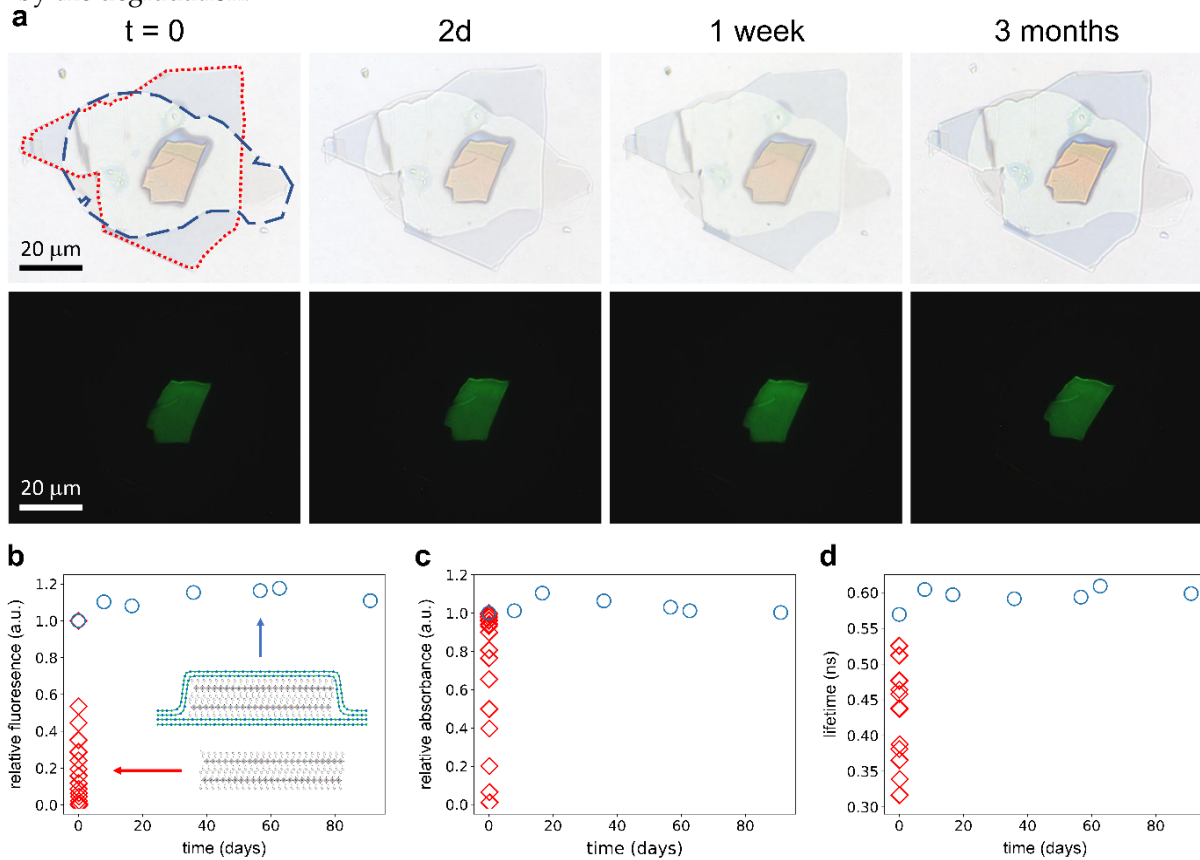

**Figure 3.** Spectral properties of a PEA$_2$PbI$_4$ 2D perovskite flake, which is encapsulated between two hexagonal boron nitride (hBN) layers. a) Transmittance (top) and fluorescence (bottom) micrographs of the encapsulated perovskite flake showing no sign of degradation after exposure to ambient conditions for over three months. In the top left micrograph, we highlight the edges of the bottom (blue dashed) and top (red dotted) hBN flake. b) Total fluorescence intensity of an encapsulated (circles) and un-encapsulated (diamonds) perovskite flake for different times under ambient condition. The intensities are normalized to the measured value at t = 0. c) Integrated absorbance (455 nm – 594 nm) of an encapsulated (circles) and un-encapsulated (diamonds) perovskite flake for different times under ambient condition. The intensities are normalized to the measured value at t = 0. d) 1/e lifetime of an encapsulated (circles) and un-encapsulated (diamonds) perovskite flake for different times under ambient condition.

To test the stability of a hBN encapsulated perovskites under light illumination, we monitor the time-dependent fluorescence intensity during exposure to laser light. We use a Picoquant 405 nm laser which is focused down on the perovskite flake creating a laser spot with a full width half max of around 1 μm. Using a laser intensity of 80 mWcm$^{-2}$, the un-encapsulated perovskite flake completely degrades within only 15 minutes as shown in **Figure 4** with a transmission photograph of the un-encapsulated flake after exposure to laser light. The local degradation through the laser is nicely visible as the degraded region (indicated with a blue arrow) does not absorb any visible light. On the other hand, the encapsulated flake shows no sign of degradation. In contrast, the fluorescence intensity even slightly increases, which could be due the photobrightening effect [67]. It is necessary to increase the laser intensity ten times to observe a sizeable drop in the fluorescence in the encapsulated flake and it is still less pronounced than the drop observed in the un-encapsulated flake with ten times lower laser intensity.

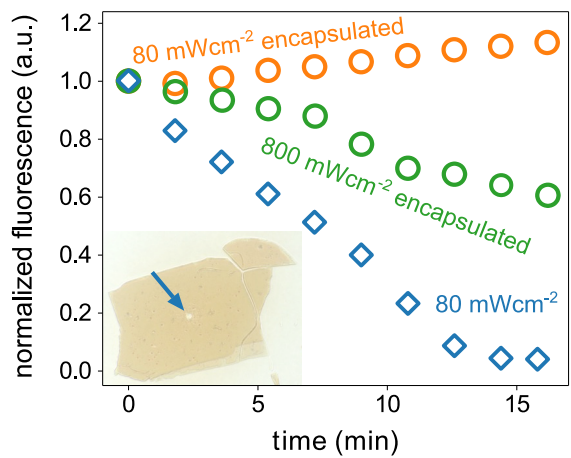

**Figure 4.** Degradation of PEA$_2$PbI$_4$ 2D perovskite flakes under 405 nm laser illumination. The graph shows the normalized (to t = 0) fluorescence intensity of encapsulated (circles) and un-encapsulated (diamonds) perovskite flakes for different times of laser exposure. The un-encapsulated flake degrades rapidly within only 15 min under 80 mWcm$^{-2}$ of laser light. For the same laser intensity, the encapsulated sample does not show any effect of degradation. When increasing the intensity tenfold to 800 mWcm$^{-2}$, also the encapsulated sample starts to degrade and decreases to around 60% of its initial brightness after 15 min of exposure. The inset shows an un-encapsulated flake (~70 μm in width) after 15 min of exposure to 80 mWcm$^{-2}$ of laser light. The flake degrades locally and becomes transparent, where the laser hit the flake (indicated by the arrow).

## 5. Conclusions

In conclusion, we have shown that it is possible to prevent the degradation under ambient conditions of PEA2PbI4 perovskite flakes through encapsulation with hBN and that hBN encapsulated flakes are stable for at least three months. Further, encapsulated flakes show an increased resistance to laser illumination. As a result, hBN encapsulation allows more thorough spectroscopic studies of the intrinsic opto-electronic properties of this material platform. Especially studies of few-layer perovskites can benefit from the increased stability, as they are particularly sensitive to environmental and light exposure. The presented encapsulation technique is not restricted to hBN but can be extended to other 2D materials that prevent the diffusion of water and air molecules, offering the possibility to have an integrated encapsulation in perovskite-2D material heterostructures.

**Supplementary Materials:** Figure S1: XRD data, Figure S2: Fluorescence and absorbance of PEA$_2$PbI$_4$, Figure S3: Logarithmic plot of data from Fig. 2b, Figure S4: Fluorescence, absorbance and lifetime traces of the encapsulated flake, Figure S5: Fluorescence and transmittance images of an un-encapsulated and partially encapsulated flake.

**Author Contributions:** Sample preparation, M.S. and P.G.; experiments and data analysis, M.S.; writing—original draft preparation, M.S. and F.P.; review and editing, M.S., P.G. A.C., and F.P.; supervision, F.P. and A.C.

**Funding:** MS acknowledges financial support of a fellowship from "la Caixa" Foundation (ID 100010434). The fellowship code is LCF/BQ/IN17/11620040. MS has received funding from the European Union´s Horizon 2020 research and innovation programme under the Marie Skłodowska-Curie grant agreement No. 713673. FP acknowledges financial support from the Spanish Ministry of Economy and Competitiveness through The "María de Maeztu" Programme for Units of Excellence in R&D (MDM-2014-0377) and the Comunidad de Madrid Talent Program for Experienced Researchers (2016-T1/IND-1209). ACG acknowledges financial support from the European Research Council (ERC) under the European Union's Horizon 2020 research and innovation programme (grant agreement n° 755655, ERC-StG 2017 project 2D-TOPSENSE) and EU Graphene Flagship funding (Grant Graphene Core 2, 785219).

**Acknowledgments:** The authors thank Dr. Beatriz H. Juárez, Diego Ruiz, and Felix Carrascoso-Plana for technical assistance.

**References**

1. Kojima, A.; Miyasaka, T.; Teshima, K.; Shirai, Y. Organometal halide perovskites as visible-light sensitizers for photovoltaic cells. *J. Am. Chem. Soc.* **2009**, *131*, 6050–6051.
2. Green, M.A.; Ho-Baillie, A.; Snaith, H.J. The emergence of perovskite solar cells. *Nat. Photonics* **2014**, *8*, 506–514.
3. Yang, S.; Fu, W.; Zhang, Z.; Chen, H.; Li, C.Z. Recent advances in perovskite solar cells: Efficiency, stability and lead-free perovskite. *J. Mater. Chem. A* **2017**, *5*, 11462–11482.
4. Green, M.A.; Hishikawa, Y.; Dunlop, E.D.; Levi, D.H.; Hohl-Ebinger, J.; Yoshita, M.; Ho-Baillie, A.W.Y. Solar cell efficiency tables (Version 53). *Prog. Photovoltaics Res. Appl.* **2018**, *27*, 2–12.
5. Congreve, D.N.; Weidman, M.C.; Seitz, M.; Paritmongkol, W.; Dahod, N.S.; Tisdale, W.A. Tunable Light-Emitting Diodes Utilizing Quantum-Confined Layered Perovskite Emitters. *ACS Photonics* **2017**, *4*, 476–481.
6. Tan, Z.-K.; Moghaddam, R.S.; Lai, M.L.; Docampo, P.; Higler, R.; Deschler, F.; Price, M.; Sadhanala, A.; Pazos, L.M.; Credgington, D.; et al. Bright light-emitting diodes based on organometal halide perovskite. *Nat. Nanotechnol.* **2014**, *9*, 687–692.
7. Pathak, S.; Sakai, N.; Wisnivesky Rocca Rivarola, F.; Stranks, S.D.; Liu, J.; Eperon, G.E.; Ducati, C.; Wojciechowski, K.; Griffiths, J.T.; Haghighirad, A.A.; et al. Perovskite Crystals for Tunable White Light Emission. *Chem. Mater.* **2015**, *27*, 8066–8075.
8. Kim, Y.H.; Cho, H.; Heo, J.H.; Kim, T.S.; Myoung, N.S.; Lee, C.L.; Im, S.H.; Lee, T.W. Multicolored organic/inorganic hybrid perovskite light-emitting diodes. *Adv. Mater.* **2015**, *27*, 1248–1254.
9. Deschler, F.; Price, M.; Pathak, S.; Klintberg, L.E.; Jarausch, D.D.; Higler, R.; Hüttner, S.; Leijtens, T.; Stranks, S.D.; Snaith, H.J.; et al. High photoluminescence efficiency and optically pumped lasing in solution-processed mixed halide perovskite semiconductors. *J. Phys. Chem. Lett.* **2014**, *5*, 1421–1426.
10. Xing, G.; Mathews, N.; Lim, S.S.; Yantara, N.; Liu, X.; Sabba, D.; Grätzel, M.; Mhaisalkar, S.; Sum, T.C. Low-temperature solution-processed wavelength-tunable perovskites for lasing. *Nat. Mater.* **2014**, *13*, 476–480.
11. Zhang, Q.; Ha, S.T.; Liu, X.; Sum, T.C.; Xiong, Q. Room-temperature near-infrared high-Q perovskite whispering-gallery planar nanolasers. *Nano Lett.* **2014**, *14*, 5995–6001.
12. Varun, S.; Stranks, S.D.; Snaith, H.J. Outshining Silicon. *Sci. Am.* **2015**, *313*, 54–59.
13. Meggiolaro, D.; Motti, S.G.; Mosconi, E.; Barker, A.J.; Ball, J.; Andrea Riccardo Perini, C.; Deschler, F.; Petrozza, A.; De Angelis, F. Iodine chemistry determines the defect tolerance of lead-halide perovskites. *Energy Environ. Sci.* **2018**, *11*, 702–713.
14. Yin, W.J.; Shi, T.; Yan, Y. Unusual defect physics in $CH_3NH_3PbI_3$ perovskite solar cell absorber. *Appl. Phys. Lett.* **2014**, *104*, 1–5.
15. Brandt, R.E.; Stevanović, V.; Ginley, D.S.; Buonassisi, T. Identifying defect-tolerant semiconductors with high minority-carrier lifetimes: Beyond hybrid lead halide perovskites. *MRS Commun.* **2015**, *5*, 265–275.
16. Steirer, K.X.; Schulz, P.; Teeter, G.; Stevanovic, V.; Yang, M.; Zhu, K.; Berry, J.J. Defect Tolerance in Methylammonium Lead Triiodide Perovskite. *ACS Energy Lett.* **2016**, *1*, 360–366.
17. Kang, J.; Wang, L.W. High Defect Tolerance in Lead Halide Perovskite $CsPbBr_3$. *J. Phys. Chem. Lett.* **2017**, *8*, 489–493.
18. Heo, J.H.; Im, S.H.; Noh, J.H.; Mandal, T.N.; Lim, C.; Chang, J.A.; Lee, Y.H.; Kim, H.; Sarkar, A.; Nazeeruddin, K.; et al. Efficient inorganic–organic hybrid heterojunction solar cells containing perovskite compound and polymeric hole conductors. *Nat. Photonics* **2013**, *7*, 486.

19. Xing, G.; Mathews, N.; Sun, S.; Lim, S.S.; Lam, Y.M.; Grätzel, M.; Mhaisalkar, S.; Sum, T.C. Long-range balanced electron- and hole-transport lengths in organic-inorganic $CH_3NH_3PbI_3$. *Science* **2013**, *342*, 344–7.
20. Stranks, S.D.; Stranks, S.D.; Eperon, G.E.; Grancini, G.; Menelaou, C.; Alcocer, M.J.P.; Leijtens, T.; Herz, L.M.; Petrozza, A.; Snaith, H.J. Electron-Hole Diffusion Lengths Exceeding. *Science* **2013**, *342*, 341–344.
21. Protesescu, L.; Yakunin, S.; Bodnarchuk, M.I.; Krieg, F.; Caputo, R.; Hendon, C.H.; Yang, R.X.; Walsh, A.; Kovalenko, M. V. Nanocrystals of Cesium Lead Halide Perovskites ($CsPbX_3$, X = Cl, Br, and I): Novel Optoelectronic Materials Showing Bright Emission with Wide Color Gamut. *Nano Lett.* **2015**, *15*, 3692–3696.
22. Weidman, M.C.; Seitz, M.; Stranks, S.D.; Tisdale, W.A. Highly Tunable Colloidal Perovskite Nanoplatelets through Variable Cation, Metal, and Halide Composition. *ACS Nano* **2016**, *10*, 7830–7839.
23. Shamsi, J.; Urban, A.S.; Imran, M.; De Trizio, L.; Manna, L. Metal Halide Perovskite Nanocrystals: Synthesis, Post-Synthesis Modifications, and Their Optical Properties. *Chem. Rev.* **2019**.
24. Niu, G.; Guo, X.; Wang, L. Review of recent progress in chemical stability of perovskite solar cells. *J. Mater. Chem. A* **2015**, *3*, 8970–8980.
25. Berhe, T.A.; Su, W.N.; Chen, C.H.; Pan, C.J.; Cheng, J.H.; Chen, H.M.; Tsai, M.C.; Chen, L.Y.; Dubale, A.A.; Hwang, B.J. Organometal halide perovskite solar cells: Degradation and stability. *Energy Environ. Sci.* **2016**, *9*, 323–356.
26. Fu, Q.; Tang, X.; Huang, B.; Hu, T.; Tan, L.; Chen, L.; Chen, Y. Recent Progress on the Long-Term Stability of Perovskite Solar Cells. *Adv. Sci.* **2018**, *5*.
27. Manser, J.S.; Saidaminov, M.I.; Christians, J.A.; Bakr, O.M.; Kamat, P. V. Making and Breaking of Lead Halide Perovskites. *Acc. Chem. Res.* **2016**, *49*, 330–338.
28. Cao, D.H.; Stoumpos, C.C.; Farha, O.K.; Hupp, J.T.; Kanatzidis, M.G. 2D Homologous Perovskites as Light-Absorbing Materials for Solar Cell Applications. *J. Am. Chem. Soc.* **2015**, *137*, 7843–7850.
29. Philippe, B.; Park, B.W.; Lindblad, R.; Oscarsson, J.; Ahmadi, S.; Johansson, E.M.J.; Rensmo, H. Chemical and electronic structure characterization of lead halide perovskites and stability behavior under different exposures-A photoelectron spectroscopy investigation. *Chem. Mater.* **2015**, *27*, 1720–1731.
30. Fang, H.H.; Yang, J.; Tao, S.; Adjokatse, S.; Kamminga, M.E.; Ye, J.; Blake, G.R.; Even, J.; Loi, M.A. Unravelling Light-Induced Degradation of Layered Perovskite Crystals and Design of Efficient Encapsulation for Improved Photostability. *Adv. Funct. Mater.* **2018**, *28*, 1800305.
31. Aristidou, N.; Sanchez-Molina, I.; Chotchuangchutchaval, T.; Brown, M.; Martinez, L.; Rath, T.; Haque, S.A. The Role of Oxygen in the Degradation of Methylammonium Lead Trihalide Perovskite Photoactive Layers. *Angew. Chemie - Int. Ed.* **2015**, *54*, 8208–8212.
32. Bryant, D.; Aristidou, N.; Pont, S.; Sanchez-Molina, I.; Chotchunangatchaval, T.; Wheeler, S.; Durrant, J.R.; Haque, S.A. Light and oxygen induced degradation limits the operational stability of methylammonium lead triiodide perovskite solar cells. *Energy Environ. Sci.* **2016**, *9*, 1655–1660.
33. Patel, J.B.; Milot, R.L.; Wright, A.D.; Herz, L.M.; Johnston, M.B. Formation Dynamics of $CH_3NH_3PbI_3$ Perovskite Following Two-Step Layer Deposition. *J. Phys. Chem. Lett.* **2016**, *7*, 96–102.
34. Leguy, A.M.A.; Hu, Y.; Campoy-Quiles, M.; Alonso, M.I.; Weber, O.J.; Azarhoosh, P.; Van Schilfgaarde, M.; Weller, M.T.; Bein, T.; Nelson, J.; et al. Reversible hydration of $CH_3NH_3PbI_3$ in films, single crystals, and solar cells. *Chem. Mater.* **2015**, *27*, 3397–3407.
35. Wei, Y.; Audebert, P.; Galmiche, L.; Lauret, J.-S.; Deleporte, E.; Wei, Y.; Audebert, P.; Galmiche, L.; Lauret, J.-S.; Deleporte, E. Photostability of 2D Organic-Inorganic Hybrid Perovskites. *Mater.* **2014**, *7*, 4789–4802.

36. Merdasa, A.; Bag, M.; Tian, Y.; Källman, E.; Dobrovolsky, A.; Scheblykin, I.G. Super-resolution luminescence microspectroscopy reveals the mechanism of photoinduced degradation in $CH_3NH_3PbI_3$ perovskite nanocrystals. *J. Phys. Chem. C* **2016**, *120*, 10711–10719.

37. Eperon, G.E.; Stranks, S.D.; Menelaou, C.; Johnston, M.B.; Herz, L.M.; Snaith, H.J. Formamidinium lead trihalide: A broadly tunable perovskite for efficient planar heterojunction solar cells. *Energy Environ. Sci.* **2014**, *7*, 982–988.

38. Kulbak, M.; Gupta, S.; Kedem, N.; Levine, I.; Bendikov, T.; Hodes, G.; Cahen, D. Cesium Enhances Long-Term Stability of Lead Bromide Perovskite-Based Solar Cells. *J. Phys. Chem. Lett.* **2016**, *7*, 167–172.

39. Saliba, M.; Matsui, T.; Domanski, K.; Seo, J.-Y.; Ummadisingu, A.; Zakeeruddin, S.M.; Correa-Baena, J.-P.; Tress, W.R.; Abate, A.; Hagfeldt, A.; et al. Incorporation of rubidium cations into perovskite solar cells improves photovoltaic performance. *Science* **2016**, *354*, 206–209.

40. Turren-Cruz, S.-H.; Hagfeldt, A.; Saliba, M. Methylammonium-free, high-performance, and stable perovskite solar cells on a planar architecture. *Science* **2018**, *362*, 449–453.

41. Lee, J.W.; Kim, D.H.; Kim, H.S.; Seo, S.W.; Cho, S.M.; Park, N.G. Formamidinium and cesium hybridization for photo- and moisture-stable perovskite solar cell. *Adv. Energy Mater.* **2015**, *5*, 1501310.

42. Saliba, M.; Matsui, T.; Seo, J.Y.; Domanski, K.; Correa-Baena, J.P.; Nazeeruddin, M.K.; Zakeeruddin, S.M.; Tress, W.; Abate, A.; Hagfeldt, A.; et al. Cesium-containing triple cation perovskite solar cells: Improved stability, reproducibility and high efficiency. *Energy Environ. Sci.* **2016**, *9*, 1989–1997.

43. Shukla, S.; Shukla, S.; Haur, L.J.; Dintakurti, S.S.H.; Han, G.; Priyadarshi, A.; Baikie, T.; Mhaisalkar, S.G.; Mathews, N. Effect of Formamidinium/Cesium Substitution and $PbI_2$ on the Long-Term Stability of Triple-Cation Perovskites. *ChemSusChem* **2017**, *10*, 3804–3809.

44. Noh, J.H.; Im, S.H.; Heo, J.H.; Mandal, T.N.; Seok, S. Il Chemical management for colorful, efficient, and stable inorganic-organic hybrid nanostructured solar cells. *Nano Lett.* **2013**, *13*, 1764–1769.

45. Misra, R.K.; Aharon, S.; Li, B.; Mogilyansky, D.; Visoly-Fisher, I.; Etgar, L.; Katz, E.A. Temperature- and component-dependent degradation of perovskite photovoltaic materials under concentrated sunlight. *J. Phys. Chem. Lett.* **2015**, *6*, 326–330.

46. Chen, Y.; Chen, T.; Dai, L. Layer-by-layer growth of $CH_3NH_3PbI_{3-x}Cl_x$ for highly efficient planar heterojunction perovskite solar cells. *Adv. Mater.* **2015**, *27*, 1053–1059.

47. Suarez, B.; Gonzalez-Pedro, V.; Ripolles, T.S.; Sanchez, R.S.; Otero, L.; Mora-Sero, I. Recombination study of combined halides (Cl, Br, I) perovskite solar cells. *J. Phys. Chem. Lett.* **2014**, *5*, 1628–1635.

48. Grancini, G.; Roldán-Carmona, C.; Zimmermann, I.; Mosconi, E.; Lee, X.; Martineau, D.; Narbey, S.; Oswald, F.; De Angelis, F.; Graetzel, M.; et al. One-Year stable perovskite solar cells by 2D/3D interface engineering. *Nat. Commun.* **2017**, *8*, 1–8.

49. Kitazawa, N.; Watanabe, Y. Preparation and stability of nanocrystalline $(C_6H_5C_2H_4NH_3)_2PbI_4$-doped PMMA films. *J. Mater. Sci.* **2002**, *37*, 4845–4848.

50. Habisreutinger, S.N.; Leijtens, T.; Eperon, G.E.; Stranks, S.D.; Nicholas, R.J.; Snaith, H.J. Carbon nanotube/polymer composites as a highly stable hole collection layer in perovskite solar cells. *Nano Lett.* **2014**, *14*, 5561–5568.

51. Bella, F.; Bella, F.; Griffini, G.; Saracco, G.; Grätzel, M.; Hagfeldt, A.; Turri, S.; Gerbaldi, C. Improving efficiency and stability of perovskite solar cells with photocurable fluoropolymers (Science, 2016, DOI: 10.1126/science.aah4046). **2016**, *4046*, 1–11.

52. Wang, D.; Wright, M.; Elumalai, N.K.; Uddin, A. Stability of perovskite solar cells. *Sol. Energy Mater. Sol. Cells* **2016**, *147*, 255–275.

53. Liao, H.; Guo, S.; Cao, S.; Wang, L.; Gao, F.; Yang, Z.; Zheng, J.; Yang, W. A General Strategy for In Situ

Growth of All-Inorganic CsPbX$_3$(X = Br, I, and Cl) Perovskite Nanocrystals in Polymer Fibers toward Significantly Enhanced Water/Thermal Stabilities. *Adv. Opt. Mater.* **2018**, *6*, 1–8.

54. Gevorgyan, S.A.; Madsen, M. V.; Dam, H.F.; Jørgensen, M.; Fell, C.J.; Anderson, K.F.; Duck, B.C.; Mescheloff, A.; Katz, E.A.; Elschner, A.; et al. Interlaboratory outdoor stability studies of flexible roll-to-roll coated organic photovoltaic modules: Stability over 10,000 h. *Sol. Energy Mater. Sol. Cells* **2013**, *116*, 187–196.

55. Island, J.O.; Steele, G.A.; Van Der Zant, H.S.J.; Castellanos-Gomez, A. Environmental instability of few-layer black phosphorus. *2D Mater.* **2015**, *2*, 011002.

56. Doganov, R.A.; O'Farrell, E.C.T.; Koenig, S.P.; Yeo, Y.; Ziletti, A.; Carvalho, A.; Campbell, D.K.; Coker, D.F.; Watanabe, K.; Taniguchi, T.; et al. Transport properties of pristine few-layer black phosphorus by van der Waals passivation in an inert atmosphere. *Nat. Commun.* **2015**, *6*, 6647.

57. Fan, Z.; Xiao, H.; Wang, Y.; Zhao, Z.; Lin, Z.; Cheng, H.C.; Lee, S.J.; Wang, G.; Feng, Z.; Goddard, W.A.; et al. Layer-by-Layer Degradation of Methylammonium Lead Tri-iodide Perovskite Microplates. *Joule* **2017**, *1*, 548–562.

58. Yu, H.; Cheng, X.; Wang, Y.; Liu, Y.; Rong, K.; Li, Z.; Wan, Y.; Gong, W.; Watanabe, K.; Taniguchi, T.; et al. Waterproof Perovskite-Hexagonal Boron Nitride Hybrid Nanolasers with Low Lasing Thresholds and High Operating Temperature. *ACS Photonics* **2018**, 4520–4528.

59. Leng, K.; Abdelwahab, I.; Verzhbitskiy, I.; Telychko, M.; Chu, L.; Fu, W.; Chi, X.; Guo, N.; Chen, Z.; Chen, Z.; et al. Molecularly thin two-dimensional hybrid perovskites with tunable optoelectronic properties due to reversible surface relaxation. *Nat. Mater.* **2018**, *17*, 908–914.

60. Yaffe, O.; Chernikov, A.; Norman, Z.M.; Zhong, Y.; Velauthapillai, A.; Van Der Zande, A.; Owen, J.S.; Heinz, T.F. Excitons in ultrathin organic-inorganic perovskite crystals. *Phys. Rev. B* **2015**, *92*, 045414.

61. Ha, S.T.; Shen, C.; Zhang, J.; Xiong, Q. Laser cooling of organic-inorganic lead halide perovskites. *Nat. Photonics* **2016**, *10*, 115–121.

62. Liang, K.; Mitzi, D.B.; Prikas, M.T. Synthesis and Characterization of Organic-Inorganic Perovskite Thin Films Prepared Using a Versatile Two-Step Dipping Technique. *Chem. Mater.* **1998**, *10*, 403–411.

63. Gauthron, K.; Lauret, J.; Doyennette, L.; Lanty, G.; Choueiry, A. Al; Zhang, S.J.; Largeau, L.; Mauguin, O.; Bloch, J.; Deleporte, E. Optical spectroscopy of two-dimensional layered (C$_6$H$_5$C$_2$H$_4$-NH$_3$)$_2$-PbI$_4$ perovskite. *Opt. Express* **2010**, *18*, 5912–5919.

64. Ma, D.; Fu, Y.; Dang, L.; Zhai, J.; Guzei, I.A.; Jin, S. Single-crystal microplates of two-dimensional organic–inorganic lead halide layered perovskites for optoelectronics. *Nano Res.* **2017**, *10*, 2117–2129.

65. Niu, W.; Eiden, A.; Vijaya Prakash, G.; Baumberg, J.J. Exfoliation of self-assembled 2D organic-inorganic perovskite semiconductors. *Appl. Phys. Lett.* **2014**, *104*, 171111.

66. Castellanos-Gomez, A.; Buscema, M.; Molenaar, R.; Singh, V.; Janssen, L.; Van Der Zant, H.S.J.; Steele, G.A. Deterministic transfer of two-dimensional materials by all-dry viscoelastic stamping. *2D Mater.* **2014**, *1*, 011002.

67. DeQuilettes, D.W.; Zhang, W.; Burlakov, V.M.; Graham, D.J.; Leijtens, T.; Osherov, A.; Bulovic, V.; Snaith, H.J.; Ginger, D.S.; Stranks, S.D. Photo-induced halide redistribution in organic–inorganic perovskite films. *Nat. Commun.* **2016**, *7*, 11683.

**SUPPORTING INFORMATION**

Figure S1a shows the x-ray diffraction (XRD) of a freshly dropcast film of PEA2PbI4 perovskite. The XRD data reveals evenly spaced peaks of 5.5° (2θ). Using Bragg's law

$$n\lambda = 2d\sin(\theta) \rightarrow \sin(\theta) = \frac{\lambda}{2d}n \qquad (1)$$

and a linear fit through the peak positions (Figure S1b), one can extract a spacing of 1.63 nm of the inorganic layers (λ = 1.54060 Å). Additionally, the absence of other diffraction peaks from the perovskite unit cell shows that the inorganic layers of the 2D perovskite grew parallel to the substrate.

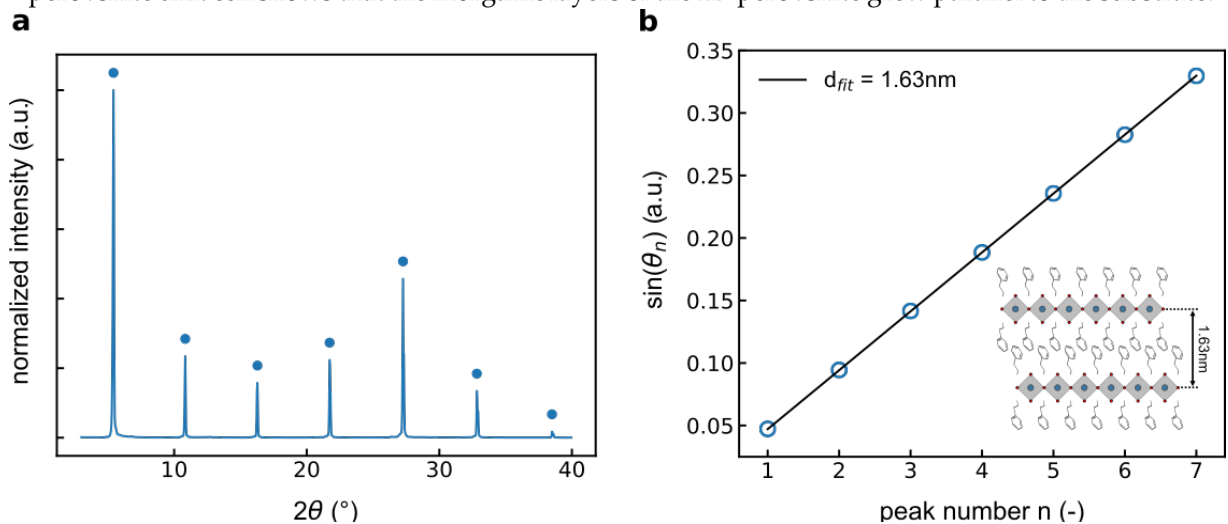

**Figure S1.** a) X-ray diffraction of PEA2PbI4 with evenly spaced peaks. b) A linear fit through the peak positions and using Bragg's law reveals a spacing of 1.63 nm of the inorganic layers. Inset shows the sketch of a 2D perovskite structure, indicating the measured spacing of the inorganic layers.

Figure S2 shows a characteristic fluorescence and absorbance spectra of an exfoliated PEA2PbI4 flake.

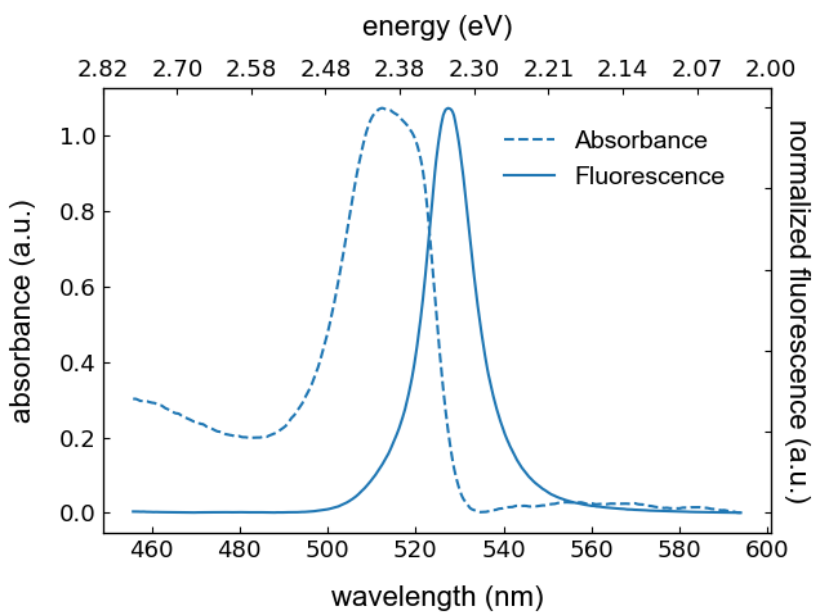

**Figure S2.** Fluorescence (solid) and absorbance (dashed) spectra of an exfoliated PEA2PbI4 2D perovskite flake.

Figure S3 shows the total normalized (to t = 0) fluorescence intensity of the un-encapsulated perovskite flake from Figure 2 of the main text. The logarithmic scale depicts three different stages. First a rapid drop to around 50% of emission intensity, which we attribute to the formation of non-radiative decay channels at the surface of the flake. Secondly, we observe an exponential decrease with a fitted decay constant of -0.78 h⁻¹, which corresponds to a decrease of 45% (= $e^{-0.78}$) per hour. We assign this second phase to the layer-by-layer degradation and the formation of additional non-radiative decay channels, which is supported by a simultaneously decreasing fluorescence lifetime (Figure 2f). Finally, the intensity decays rapidly to zero which we assign to the complete degradation of the sample through the lateral degradation which is faster than the layer-by-layer degradation (we measure the fluorescence only in the center of the flake, which explains why the lateral degradation has no significant influence in the first two steps).

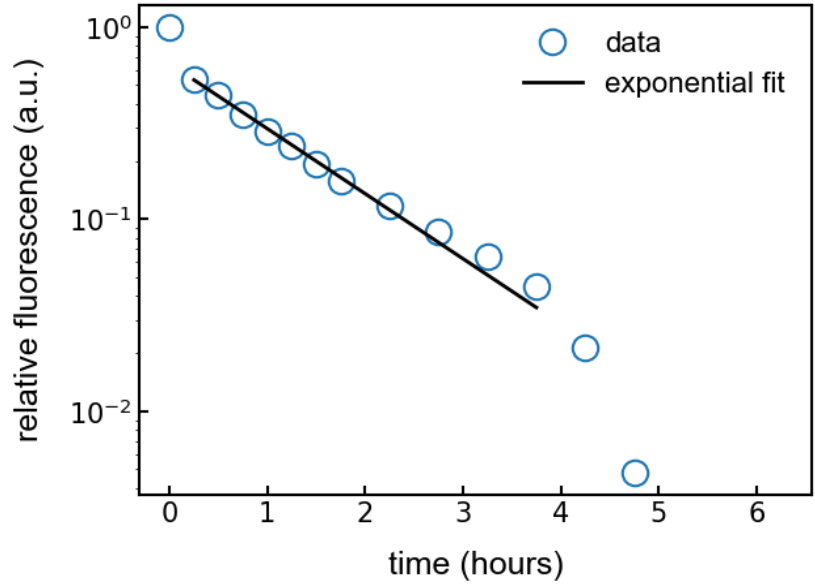

**Figure S3.** Total normalized (to t = 0) fluorescence intensity of an un-encapsulated perovskite flake under ambient conditions. Data shown here is the same as in Figure 2b of the main text, but with a logarithmic ordinate.

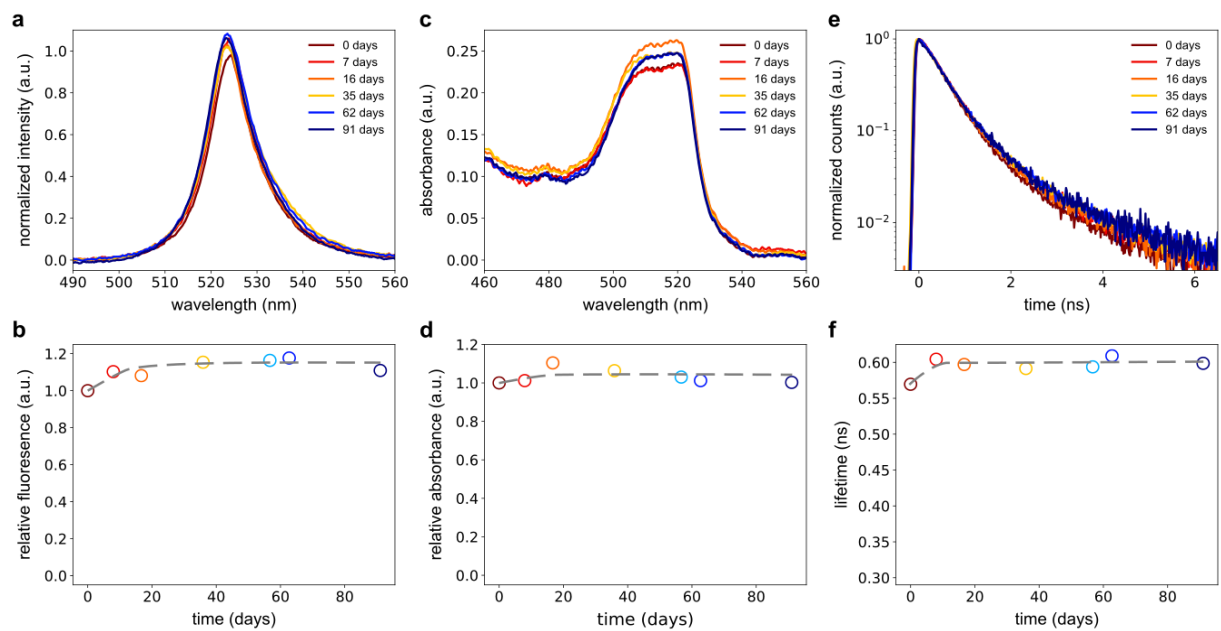

**Figure S4.** Same data as in Figure 3b-d, but including the traces. Spectral properties of the encapsulated PEA$_2$PbI$_4$ 2D perovskite flake shown in Figure 3 of the main text for different times under ambient exposure. a, c, e) Show the fluorescence, absorbance and fluorescence lifetime traces, respectively, for different times of ambient exposure. b) Total fluorescence intensity from (a) normalized to the measured intensity at t = 0. d) Integrated absorbance (455 nm – 594 nm) from (c) and normalized to the measured absorbance at t = 0. f) Extracted 1/e lifetime from the fluorescence lifetime traces in (e). Grey lines in b), d), and f) are guides to the eye.

Figure S5 shows an encapsulated and an un-encapsulated perovskite flake for different times under ambient exposure. While the un-encapsulated flake readily degrades within one day, the hBN encapsulated flake is stable for a much longer time. However, the hBN seal of the encapsulated flake has a leak in the seal (indicated by an arrow), which allows air molecules to slowly diffuse towards the perovskite flake. Nevertheless, it is interesting to note that the center of the flake remains mainly unaffected by the degradation even after 1 month, underlining that the main purpose of the hBN seal is to shield the perovskite from air molecules.

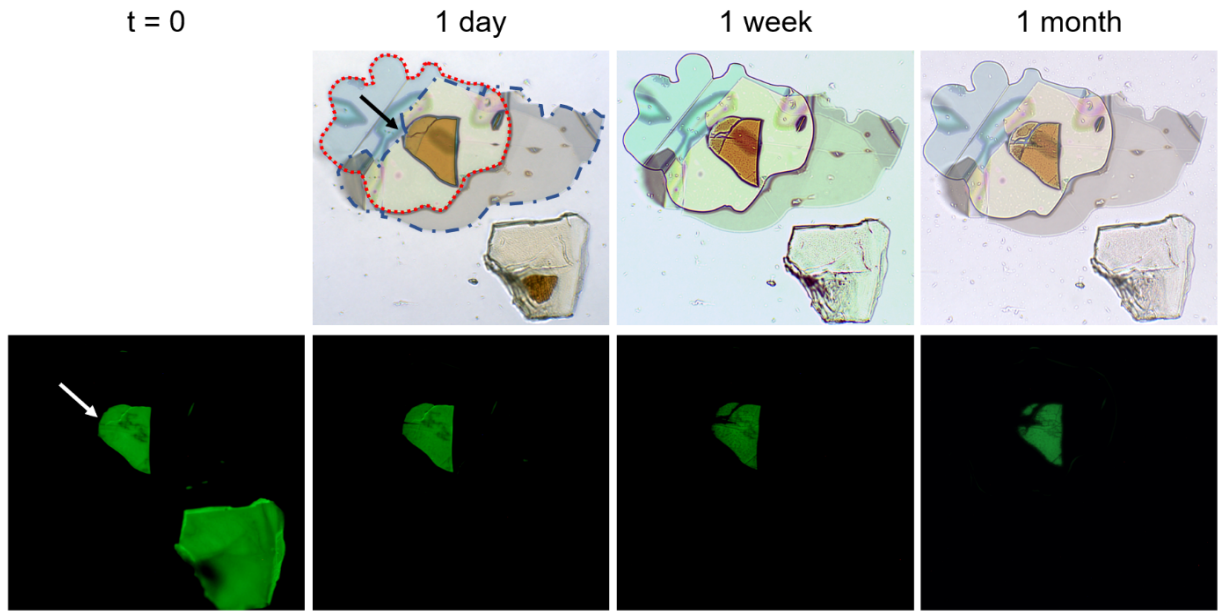

**Figure S5.** Transmission (top row) and fluorescence (bottom row) micrographs of an un-encapsulated (bottom right of each micrograph) and partially encapsulated (top left of each micrograph) perovskite flake for different times under ambient conditions. The four columns of the figure correspond to 0, 1 day, 1 week, and 1 month of ambient exposure. The un-encapsulated flake degrades within one day, while the partially encapsulated flake is stable for a much longer time. However, the hBN seal of the partially encapsulated flake has a leak in the seal (indicated by an arrow), which allows air molecules to slowly diffuse towards the perovskite flake leading to the degradation of the flake which spreads from the leak inwards.